\documentclass{egpubl}
\usepackage{eg2019}

\ConferencePaper

\electronicVersion

\ifpdf \usepackage[pdftex]{graphicx} \pdfcompresslevel=9
\else \usepackage[dvips]{graphicx} \fi

\PrintedOrElectronic

\usepackage{t1enc,dfadobe}
\usepackage{egweblnk}
\usepackage{cite}
\usepackage[export]{adjustbox}
\usepackage{booktabs} %
\usepackage{enumitem}
\usepackage[ruled]{algorithm2e} %
\usepackage{svg}
\usepackage{cleveref}
\crefname{enumi}{guideline}{guidelines}
\Crefname{enumi}{Guideline}{Guidelines}
\usepackage{pifont}
\usepackage{dblfloatfix}
\usepackage{amsmath}

\title[Puppet Dubbing]%
      {Puppet Dubbing}

\author[O. Fried \& M. Agrawala]
{\parbox{\textwidth}{\centering Ohad Fried$^1$ and Maneesh Agrawala$^1$}
        \\
{\parbox{\textwidth}{\centering $^1$Department of Computer Science\\
         Stanford University
        }
}
}

\newcommand*{\ShowNotes}{}
\newcommand{\ignorethis}[1]{}

\newcommand{\Reals      }     {{\textrm{I\kern-0.18em R}}}

\newcommand{\change     } [1] {\mbox{{\footnotesize $\Delta$} \kern-3pt}#1}

\newcommand{\Abs        } [1] {{\left| #1 \right|}}
\newcommand{\norm       } [1] {{\| #1 \|}}

\newlength{\w}

\definecolor{darkred}{rgb}{0.7,0.1,0.1}
\definecolor{darkgreen}{rgb}{0.1,0.5,0.1}
\definecolor{cyan}{rgb}{0.7,0.0,0.7}
\definecolor{dblue}{rgb}{0.2,0.2,0.8}
\definecolor{maroon}{rgb}{0.76,.13,.28}
\definecolor{burntorange}{rgb}{0.81,.33,0}

\ifdefined\ShowNotes
  \newcommand{\colornote}[3]{{\color{#1}\bf{#2 #3}\normalfont}}
\else
  \newcommand{\colornote}[3]{}
\fi

\newcommand {\red}[1]{\color{red}{#1}}
\newcommand {\blue}[1]{\color{blue}{#1}}

\newcommand\todosilent[1]{}

\newcommand{\cmark}{\textcolor{darkgreen}{\ding{51}}}%
\newcommand{\xmark}{\textcolor{darkred}{\ding{55}}}%

\begin{document}

\teaser{
  \setlength{\w}{\linewidth}
  \centering
  \includegraphics[width=\w]{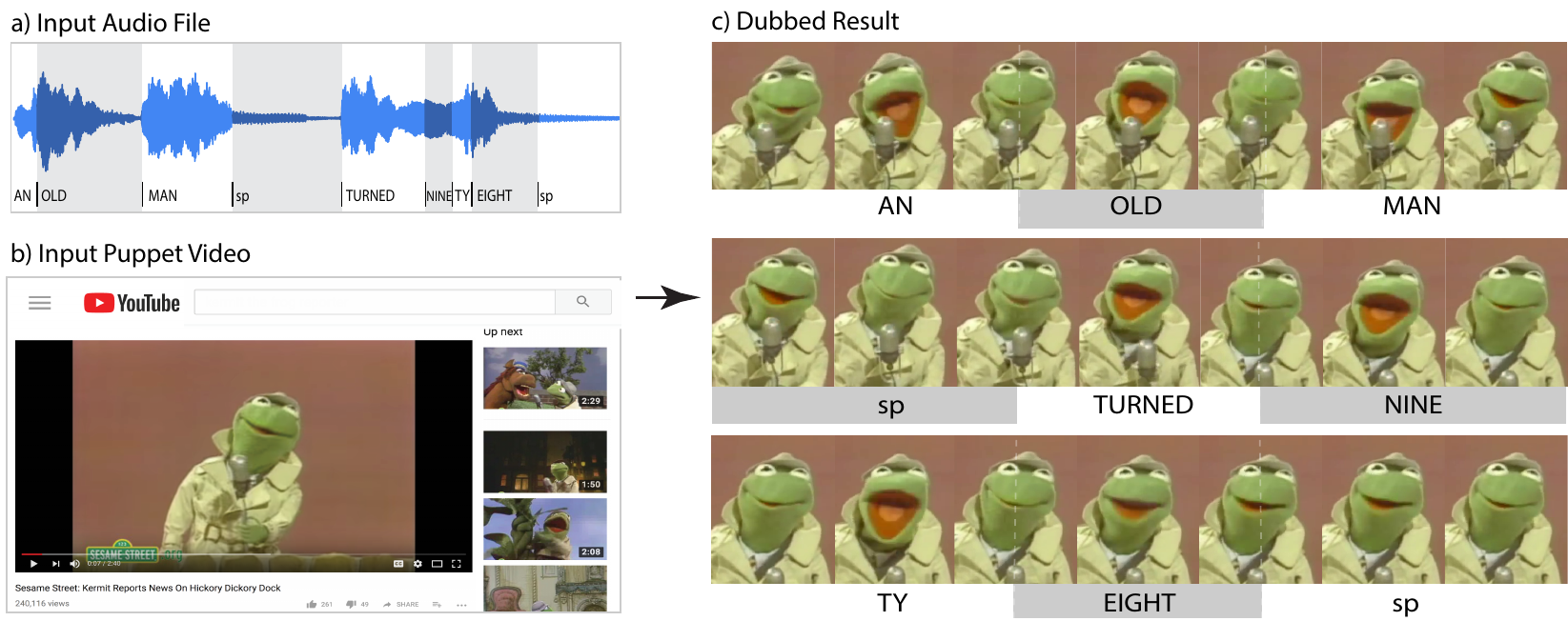}
  \caption{Given an audio file and a puppet video, we produce a dubbed result in which the puppet is saying the new audio phrase with proper mouth articulation. Specifically, each syllable of the input audio matches a closed-open-closed mouth sequence in our dubbed result. We present two methods, one semi-automatic appearance-based and one fully automatic audio-based, to create convincing dubs.
  }
  \label{fig:teaser}
}
 
\maketitle

\begin{abstract}
Dubbing puppet videos to make the characters (e.g. Kermit the Frog) convincingly speak a 
new speech track is a popular activity with many examples of
well-known puppets speaking lines from films or singing rap songs.
But manually aligning puppet mouth movements to match a new speech
track is tedious as each syllable of the speech must match a
closed-open-closed segment of mouth movement for the dub to be
convincing.
In this work, we present two methods to align a new speech track with
puppet video, one semi-automatic {\em appearance-based} and the other
fully-automatic {\em audio-based}.
The methods offer complementary advantages and disadvantages. Our
appearance-based approach directly identifies closed-open-closed
segments in the puppet video and is robust to low-quality audio as
well as misalignments between the mouth movements and speech in the
original performance, but requires some manual annotation. Our
audio-based approach assumes the original performance matches a
closed-open-closed mouth segment to each syllable of the original
speech. It is fully automatic, robust to visual occlusions
and fast puppet movements, but does not handle misalignments in the
original performance. 
We compare the methods and 
show that both improve the credibility of the resulting video over
simple baseline techniques, via quantitative evaluation and user ratings.
\end{abstract}
 
\clearpage

% ---------------------------------------------------------------------------

\section{Introduction}
\label{sec:intro}

Puppet-based video is widely available online at sites like YouTube
and Vimeo in the form of clips from well-known TV shows and films
(e.g. {\em Sesame Street, Barney and Friends, The Muppet Movie,} etc.).  The
abundance of such clips has led to a vibrant remix culture in which
people dub the clips to alter the speech and make the
puppets tell jokes\,\cite{ytPuppetJoke2014}, recite lines from famous films\,\cite{ytPuppetTarantino2014,ytPuppetPesciDeniro2008},
speak in a different language or accent\,\cite{ytPuppetItalianDub2014,ytPuppetKermitJamaican2014} or sing rap songs\,\cite{ytPuppetWarrenG2016}.

But, dubbing such puppet video is challenging as it requires carefully
matching the mouth movements of the puppet to a new speech
track. Expert puppeteers suggest that puppets are most convincing when
each closed-open-closed segment of the puppet mouth corresponds to
exactly one syllable of speech~\cite{currell1987}. 
So, the best dubbing efforts
usually involve frame-level matching and re-timing of
closed-open-closed mouth segments in the video to the syllables in the
new speech. Today, such matching and re-timing is largely a manual
process and is extremely time-consuming.

In this work, we present two techniques that significantly reduce (or
eliminate) the manual effort required to dub a puppet video with a new
source speech track (\Cref{fig:teaser}).
Both methods start by breaking the new speech into a sequence of syllables.
Our {\em appearance-based} method tracks the puppet head and lip motion to
identify the closed-open-closed segments.
We call these segments {\em visual
  syllables} and align them to the syllables in the new speech.
Our {\em audio-based} method assumes that the original puppeteer properly
aligned the visual syllables to the original puppet speech. It
treats the audio syllables in the original puppet speech as a proxy for the
visual syllable boundaries and aligns them to the syllables in the new speech.
Both methods then re-time the video and the new speech to best match each other.

Our puppet dubbing methods offer different strengths and
weaknesses. The appearance-based method is robust to low quality audio
and to the presence of background music in the original video, while
the audio-based method is robust to low visual quality and 
occlusions in the video. The appearance-based method does not assume that the
original puppeteering performance was well synced with the audio, and
can distinguish between on- and off-camera speech, but at a cost of
some manual annotation. The audio-based method requires no annotation,
but does not directly identify visual syllables and therefore can
produce artifacts when the original speech is misaligned with the
original puppet video.

The main contributions of this work are:
\begin{itemize}
  \item Identification of puppeteering guidelines that produce a convincing puppet performance.
  \item Instantiaion of these guidelines in two novel methods for dubbing existing puppet video with new audio.
  \item A novel approach for combined video and audio re-timing, which takes into account the rate of change in audio speed.
\end{itemize}

We quantitatively measure the accuracy of our puppet-specific
visual detectors and the amount of stretching and squeezing performed on the
videos, which we try to minimize to reduce artifacts.
Finally we conduct a user study which finds that our appearance-based
and audio-based methods are seen as having significantly higher
quality than simple baseline approaches.

% ---------------------------------------------------------------------------

\section{Related Work}
\label{sec:related}

Our puppet dubbing methods build on three areas of related work.

\vspace{0.05in}
\noindent
\textbf{\emph{Dubbing humans.}}
Prior work on dubbing has primarily focused on synthesizing video or
3D animation of human faces that match an input audio speech
track. These techniques fall into three main categories. (1)
Phoneme-based approaches extract the sequence of phonemes in the input
speech, generate the visual counterpart of lip and face motion for
each phoneme (called a viseme), and concatenate the visemes to form
the final result~\cite{bregler1997,edwards2016,taylor2012,wang2012}.
(2) Machine-learning approaches learn the mapping between low-level
features of the input speech to the output frames of video or 3D
animation~\cite{brand1999,karras2017,suwajanakorn2017,taylor2017} (3)
Performance capture approaches require an input RGB or RGB-D video of
an actor performing the speech and re-map the performance onto the
output video, 3D animation or even a single
image~\cite{doi:10.1111/cgf.12552,kim2018DeepVideo,Furukawa:2016:VRA:2945078.2945097,thies2016,weise2011,fyffe2014,beeler2011,cao2015,averbuch2017}.
Because human viewers are well attuned to the nuances of human faces,
it is a significant challenge for these approaches to generate
believable lip and face motions that avoid the uncanny valley.
In contrast, viewers are more forgiving with non-human
puppets and, as we will show, there are only a few key
characteristics of puppet mouth motion (as opposed to human mouth
motion) that must be maintained to produce a convincing dub. %

\vspace{0.05in}
\noindent
\textbf{\emph{Dubbing non-human characters.}}
Researchers have used some of the
performance capture techniques mentioned
earlier~\cite{weise2011,fyffe2014,beeler2011,cao2015,characterAnim} to transfer
human performances of speech to non-human characters (either 2D
cartoons or 3D models). Others have developed techniques to drive
mouth motions of such characters using audio peak
detection~\cite{barnes2008} or a phoneme-based analysis of the
speech~\cite{edwards2016}. Unlike our approach, these methods all
require rigged models of the mouth and lips of the 2D/3D characters.

\vspace{0.05in}
\noindent
\textbf{\emph{Speech-driven audio/video editing.}}
Researchers have recently developed a number of tools for
transcript-driven editing of speech-based audio and video content.  These
tools rely on an accurate text transcript of the input speech, either via
speech-to-text tools~\cite{ibmwatson} or crowdsourcing services (e.g. \url{rev.com}), and phoneme-level time
alignment of the transcript to the speech~\cite{rubin2013,gentle}. They enable text-based editing of talking-head
style video~\cite{berthouzoz2012} and
podcasts~\cite{rubin2013,shin2016}, vocally annotating
video~\cite{pavel2016,truong2016}, indexing lecture~\cite{pavel2014}
and blackboard video~\cite{shin2015}, synthesis of new speech audio to
stylistically blend into the input speech~\cite{jin2017} and automatic
editing of rough cuts~\cite{leake2017}.
We similarly rely on time-aligned text transcripts of the speech in
both the input audio and candidate video. However, instead of editing,
we focus on using the transcribed syllables for dubbing.

\section{Guidelines for Performing Puppet Speech}
\label{sec:guidelines}

Our puppet dubbing approach is inspired by puppeteering tutorials and
observations of expert puppeteer performances.  Based on these
observations we have identified three main guidelines for producing
visually convincing puppet speech:

\begin{enumerate}[label=(g\arabic*),leftmargin=.7cm]
\item \label{gline1} Each syllable in speech should match to one
  closed-open-closed segment of puppet lip motions. We call such video segments {\em visual syllables}.
\item \label{gline2} Lips should be still and ideally closed when the puppet is not speaking as
  lip motions during silences can be disturbing. We call silences in speech {\em silence syllables} and
  closed mouth video segments {\em visual silence syllables}.
\item \label{gline3} In rapid speech sequences, instead of a
    one-to-one match, several spoken syllables may correspond to a
    single visual syllable.
\end{enumerate}

\Cref{gline1,gline2} are mentioned in puppeteering training 
videos\,\cite{ytPuppetLipSync2010}.  Although \ref{gline3} is less
directly documented, we often observed a many-to-one match between
multiple syllables in fast speech and a single visual syllable, even
in expert performances. 
This is at times a conscious effort by the puppeteer to simplify the performance, or due to real-world operating
constraints on the puppets, and is related to the loss of fricatives
and elision of vowels in rapid human speech~\cite{jones2011}.
In the remainder of this paper we will use the term {\em syllable} to
refer to both silence and non-silence speech syllables and use the
specific terms only when necessary for disambiguating between the two
types.  Similarly, we will use the term {\em visual syllable} to refer
to both silence and non-silence visual syllables. 

\begin{figure}[t]
  \setlength{\w}{\columnwidth}

  \centering
  
  \includegraphics[width=\w]{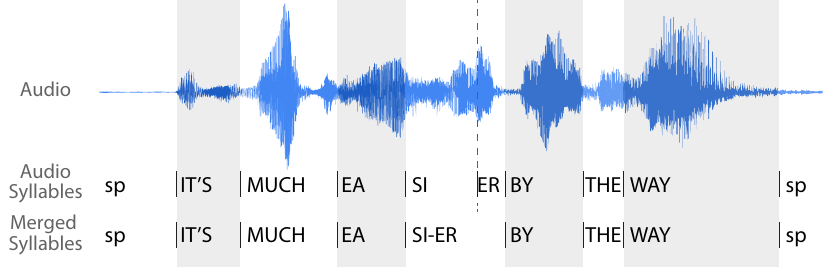}
  
  \caption{Automatic audio alignment for the phrase ``it's much easier by the way''. Given audio with transcript we locate words and phonemes, and derive syllables from them. Each resulting syllable $a_i$ is comprised of a
    label $a_i^{lbl}$ such as IT'S, MUCH and EA. The label {\em sp} indicates silence. Each syllable also includes a start time $a_i^{in}$ and end time $a_i^{out}$ indicated by the syllable segment end points.
    We merge short syllables with their neighbors to produce a final syllable segmentation.
  }
  \label{fig:syl_alignment}
\end{figure}
 
\section{Algorithmic Methods}
\label{sec:methods}

Our goal is to dub a given puppet video $V$ with new speech audio $A$
according to the guidelines in Section~\ref{sec:guidelines}.  We
assume that the new audio track is shorter than the video track.  Our
approach involves four main steps. (1) We segment the new speech audio
track into a sequence of syllables (Section~\ref{sec:audioSeg}). (2) We
segment the puppet video into a sequence of visual syllables (Section~\ref{sec:videoSeg}).  (3) We align the audio
syllables with the optimal subsequence of visual syllables
(Section~\ref{sec:alignment}). (4) Finally, we re-time the audio and video so that each
visual syllable matches the length of a corresponding audio syllable
(Section~\ref{sec:retiming}).

\subsection{Step 1: Segment Audio Into Syllables}
\label{sec:audioSeg}

To segment the new speech audio $A$ into syllables, we first obtain a
transcript of it. In practice most of our examples use the closed
captions from YouTube, but we have also experimented with automatic
speech transcription tools~\cite{ibmwatson,gentle} and crowdsourcing
transcription services like \texttt{\url{rev.com}}. We align the transcript to
the audio using P2FA~\cite{yuan2008,rubin2013}, a phoneme-based alignment
tool, and then combine the phonemes into syllables using the approach
of Bartlett et al.~\shortcite{bartlett2009}. This gives us an ordered
sequence $A = (a_1, \ldots, a_n)$ of syllables, each with a label
denoting the syllable name, start time and end time $a_i =
(a_i^\textit{lbl}, a_i^\textit{in}, a_i^\textit{out})$ (\Cref{fig:syl_alignment}).
The syllable label {\em sp} indicates a silence syllable.

While the resulting syllable to speech alignment is usually very good,
background music, environmental noise and/or poor enunciation can
create some misalignments.  Users can optionally fix such
misalignments using PRAAT~\cite{boersma2001} to adjust syllable
boundaries. In practice, we've found that 1 minute of speech
requires between 0 and 20 minutes of manual tweaking to produce a perfect
alignment, and usually less than 5 minutes. However, even with misalignments introduced by P2FA our
results are often acceptable.  We present results with and without
these manual tweaks [supplemental material: audio alignment].

\Cref{gline3} suggests that in rapid speech, puppeteers do not articulate every short syllable and instead merge them into a single visual syllable.
Thus, if a syllable is shorter than a threshold (set empirically to 150ms) we merge it with its shortest neighboring syllable. We merge in ascending length order, until no more merges are possible.
Importantly, we only merge syllables belonging
to the same word and not across words, according to puppeteering
practices, as the latter produces visible artifacts.

\begin{figure}[t]
  \setlength{\w}{\columnwidth}
  
  \centering
  
  \includegraphics[width=\w]{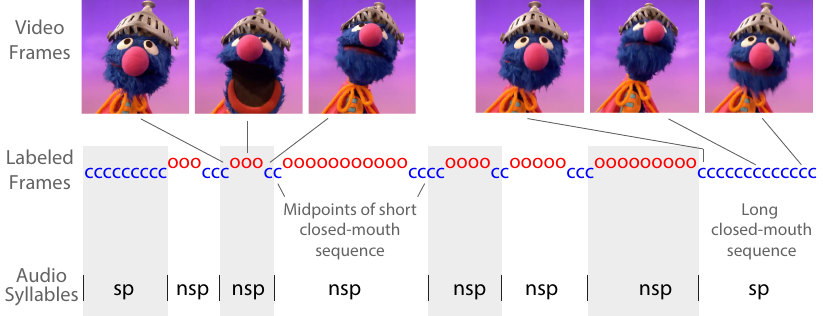}
 \caption{Converting a sequence of \textit{{\red o}pen-mouth} (red), \textit{{\blue c}losed-mouth} (blue), and \textit{invalid}
    labels into visual syllables.  A sequence of five or more
    consecutive \emph{closed-mouth} frames is deemed a silent visual
    syllable. All the remaining valid frames are grouped into
    non-silent visual syllables. Boundaries between non-silent visual
    syllables are set in the middle of the remaining short
    \emph{closed-mouth} sequences. The sequence shown contains two
    silent syllables (\textit{sp}) and six non-silent syllables (\textit{nsp}).
  }
  \label{fig:algo_sp_nsp}
\end{figure}
 
\subsection{Step 2: Segment Video Into Visual Syllables}
\label{sec:videoSeg}

We offer two methods for segmenting the puppet video $V$ into a
sequence of visual syllables $V = (v_1, \ldots, v_m)$, one that is
appearance-based and one that is audio-based.  Like audio syllables,
each resulting visual syllable consists of a label, in this case
denoting whether the syllable is silent {\em sp} or non-silent {\em
  nsp}, a start time and an end time $v_i = (v_i^\textit{lbl},
v_i^\textit{in}, v_i^\textit{out})$.  We describe each segmentation
technique and discuss their different strengths and weaknesses

\subsubsection{Appearance-based visual syllables.}
\label{subsubsec:appearance-based}

Our appearance-based algorithm is designed to first classify each frame into one of three categories, \textit{open-mouth}, \textit{closed-mouth} and \textit{invalid} as described below and then use the classification to construct visual syllables as follows.  
Given the per-frame classification (middle row of \Cref{fig:algo_sp_nsp}), we mark a sequence of five or more consecutive \emph{closed-mouth} frames as a silent visual syllable. All the remaining valid frames are grouped into non-silent visual syllables. We set the boundaries between non-silent visual syllables in the middle of the adjacent short \emph{closed-mouth} sequences. Consecutive invalid frames are grouped into invalid sequences, and are not used in later steps (their matching cost is set to infinity).

We created an annotation UI that allows quick keyboard-based annotation.
We provide hotkeys for the 3 labels and for frame navigation, requiring one keystroke per annotated label while allowing the annotator to revisit and change previous labels. The average annotation rate using our UI is 200 frames per minute.
Thus, for a 1 minute video running at 30fps annotating the complete video would take about 9 minutes.
While it is possible to use this interface to manually label short
videos, for longer videos we have developed a machine learning
approach that significantly lightens the annotation workload.

While researchers have
developed a number of facial landmark point detectors for human
faces\,\cite{ranjan2017,saragih2009} these methods are not designed to
handle puppets.   
Therefore, for each puppet video $V$ we train a mouth state
classifier to detect \emph{open-mouth}, \emph{closed-mouth} and \emph{invalid} frames. Invalid frames are those without a visible puppet head.

\begin{figure}[t]
  \setlength{\w}{\columnwidth}
  
  \centering
  
  \includegraphics[width=\w]{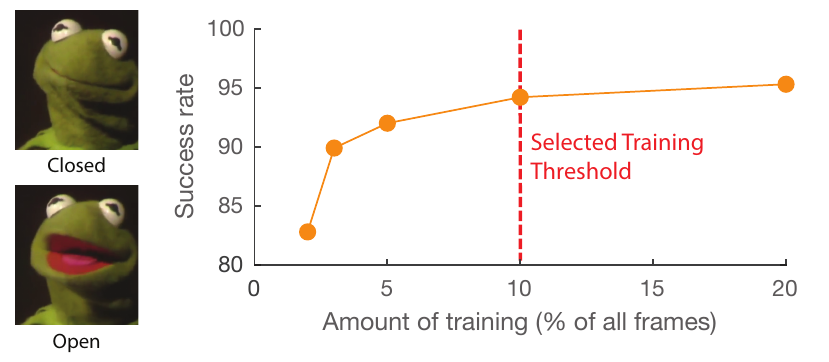}
  
  \caption{For appearance-based visual syllables we annotate open and closed mouth frames (left). We also annotate frames with no puppets as invalid (not shown). 
  For a given video we annotate a fraction of the frames and train a neural network to predict missing labels. We experimented with different annotation amounts 
and found that 10\% annotation provides a good trade-off between prediction accuracy and annotation time (right). 
}
  \label{fig:eval_detection_result}
\end{figure}
 
Specifically, we start with the pre-trained
GoogLeNet\,\cite{Szegedy_2015_CVPR} model and specialize it to our
puppet mouth state detection task. We remove the last three layers of
GoogLeNet, which are specific to the original classification task,
and replace them with a fully connected layer followed by softmax and
a classification output layer that generates one of our three
labels.
To reduce training time we focus on primarily learning the weights for the new layer; we set the
learning rate of the first 110 layers to $0$, all other
pre-existing layers to $10^{-4}$ and the new fully connected layer to $10^{-3}$.

For training data, we manually annotate a fraction of the frames for a given input video. There is a classification quality vs. annotation effort trade-off in selecting the annotation amount. 
\Cref{fig:eval_detection_result} shows the relationship between the two for a typical video. We find that annotating 10\% of the frames is a good threshold allowing a tenfold decrease in annotation time, while maintaining 94\% classification accuracy.
Using our annotation interface, manually labeling 10\% of randomly chosen frames for a 1 minute video (at 30fps) requires 0.9 minutes of annotation time --- less than the time it takes to watch the video.
We augment the training data by adding random translations and flips, and train for 6 epochs using a batch size of 10.
See [supplemental materials: 10\% vs. 100\% annotation]
for results using fully annotated and partially annotated videos.
For many examples the results are comparable, with errors in classification manifesting as consecutive visual syllables being detected as one, or one syllable being split into several.

As an optional step, users can correct errors in the classification
results using our annotation interface. We find that in practice
correcting errors requires watching the video annotated with the
labels output by our classifier, pausing it whenever an error is
spotted, stepping back a few frames and relabeling around the
error. Such correction requires an additional annotation time of about
2 minutes per 1 minute of video, but improves the labels, making them
indistinguishable from manual annotation of the complete video.

\subsubsection{Audio-based visual syllables.}
\label{subsubsec:audio-based}

Our audio-based algorithm assumes that visual
syllables in $V$ are closely aligned with the original speech audio
track in $V$. Therefore we can segment the video into visual syllables
by first segmenting its original audio track using the approach
described in Section~\ref{sec:audioSeg}. We then treat the resulting
audio syllables as a proxy for the visual syllables of the video.

\subsubsection{Comparison.}

\Cref{tbl:comparison} compares our appearance- and audio-based
visual syllable extraction approaches. 
Appearance-based visual syllables 
are robust to noisy audio
and do not rely on good audio-video alignment in the original
performance. Appearance-based visual syllables also distinguish
between on- and off-camera speech (i.e. if a character is turned away
from the camera and speaking, appearance-based syllables will not use
it as an audible syllable).
Audio-based visual syllables
use audio as a
proxy for the location of visual syllables.
They are robust to low quality video and occlusions and do not require any manual annotation.
As we will discuss in Section~\ref{sec:discussion}, 
The complementary
properties of these two approaches make each of them best suited
to different dubbing tasks. 

\begin{table}[t!]

  \newcommand{\linesl}[1]{\begin{tabular}{@{}l@{}}#1\end{tabular}}

  \centering
  \small{
  \begin{tabular}{@{}lccc}
    \toprule
                                                            & \textbf{Apearance}    & \textbf{Apearance with}     & \textbf{Audio}   \\
                                                            &                       & \textbf{manual correction}  & \\
    \midrule
    \linesl{Mean annotation time}                           & 0.9x                  & 3x                          & None             \\[0.12cm]
    \linesl{Robust to noisy audio \\ \& background music}   & \cmark                & \cmark                      & \xmark           \\[0.3cm]
    \linesl{Robust to audio-video \\ misalignment}          & \cmark                & \cmark                      & \xmark           \\[0.3cm]
    \linesl{Distinguishes on- and \\ off-camera speech}     & \cmark                & \cmark                      & \xmark           \\[0.3cm]
    \linesl{Robust to low-quality \\ video \& occlusions}   & \xmark                & \xmark                      & \cmark           \\
    \bottomrule
  \end{tabular}
  }
  \caption{Comparison between appearance-based and audio-based visual syllables. Annotation time (averaged across annotators and media assets) compares between length of media and annotation time. E.g. 3x is 3 minutes annotation time per 1 minute of video.}
  \label{tbl:comparison}
\end{table}
 
\subsection{Step 3: Align Audio Syllables to Visual Syllables}
\label{sec:alignment}

\begin{figure}[t]
  \setlength{\w}{\textwidth}
  
  \centering
  
  \includegraphics[width=\columnwidth]{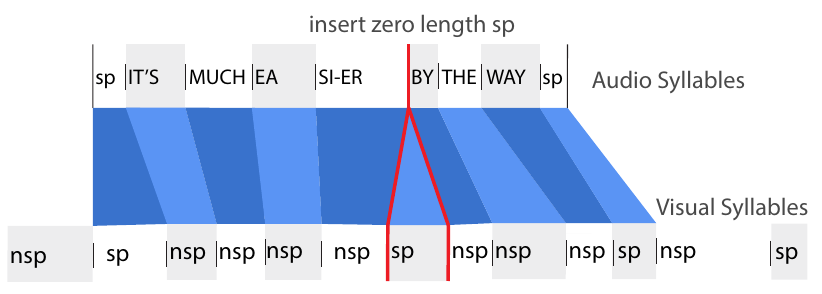}
  
  \caption{Alignment. For a sequence of audio syllables $(a_1, \ldots, a_n)$ (top) we find the best starting point within the visual syllable $(v_1, \ldots, v_m)$ (bottom) such that silence syllables align as much as possible and the difference in syllable lengths is minimized. After finding this starting point we make sure both sequences have the same number of syllables by adding zero length silence syllables opposite any unmatched silence (marked in red).}
  \label{fig:matching}
\end{figure}
 
The goal of the alignment step is to align the speech syllables to an optimal sequence of visual syllables in the puppet video, such that when possible (1) silence matches to silence, (2) non-silence matches to non-silence and (3) syllable lengths are as similar as possible. Conditions (1) and (2) ensure that when the new speech track is silent, the puppet's mouth is closed, and when there is speech, the puppet's mouth is moving. Condition (3) minimizes the amount of retiming required in Step 4 to match the timing of the visual syllables to the speech, thereby reducing artifacts.

Formally, given $A$, $V$ and their respective starting
indices $i$, $j$, we can recursively define the distance
$\mathcal{D}(A_i, V_j)$ between the sequences (Equation~\ref{eqn:dist_func}),
using $A_i=(a_i\ldots a_n)$ and $V_j=(v_j \ldots v_m)$ as shorthand for the syllable subsequences starting at $i$ and $j$ respectively.   
{
\begin{equation}
\scriptstyle
\mathcal{D}(A_i, V_j) =
\begin{cases}{\label{eqn:dist_func}}
\mathcal{D}(A_{i+1}, V_{j+1})                                  & a_i = sp, v_j = sp         \\ 
\Abs{\norm{a_i} - \norm{v_j}} + \mathcal{D}(A_{i+1}, V_{j+1})  & a_i \neq sp, v_j \neq sp   \\
w * \norm{a_i} + \mathcal{D}(A_{i+1}, V_j)                     & a_i = sp, v_j \neq sp      \\
w * \norm{v_j} + \mathcal{D}(A_i, V_{j+1})                     & a_i \neq sp, v_j = sp     
\end{cases}
\end{equation}
}
We use $\norm{a}$ to indicate the length of syllable $a$. This equation is defined for $i \le \Abs{A}$ and $j \le \Abs{V}$, and equals 0 when $i > \Abs{A}$ or $j > \Abs{V}$. 
The first term incurs no penalty when speech silence matches with visual silence, in accordance with condition (1). The second term penalizes the difference in length of the non-silent syllables, in accordance with conditions (2) and (3).  
The third and fourth terms heavily penalize alignment of silence to non-silence by adding $w$ times the length of silence. We empirically set $w$ to 1. 
Notice that in these two terms, only the sequence containing the silent syllable advances.
Therefore, we terminate this recursion only when the distance considers an equal number of non-silent syllables in the speech and audio.
Equation~\ref{eqn:dist_func} can be interpreted as a variant of dynamic time warping \cite{1163055}, in which insertions and deletions are only permitted for specific types of syllables, and both syllable length and type are used to determine the cost of operations. 

We iterate over all possible starting points of $V$ to find the best
starting point $j$, for which $\mathcal{D}(A_1, V_j)$ is minimal.  The
starting point 
defines a video sub-sequence $(v_j, \dots, v_k)$ which
best matches the audio $(a_1, \dots, a_n)$ (\Cref{fig:matching}). These
sequences have the same number of non-silent syllables, but may
include a different number of silences.  For ease of annotation we
equalize the number of syllables.  Beginning at the starting point, we
jointly iterate the two sequences. Whenever one sequence contains silence
and the other contains a non-silence, we add a zero length sized
silent syllable to produce a matching silent pair. After this syllable equalization, we are left with syllable sequences $(a_1, \dots, a_n)$ and $(v_j, \dots, v_{j+n-1})$ where $a_1$ matches $v_j$, $a_2$ matches $v_j+1$, etc.
 
\subsection{Step 4: Re-time Audio and Video}
\label{sec:retiming}

Given the source audio and target video, along with their matching
syllable sequences $(a_1, \dots, a_n)$, $(v_j, \dots, v_{j+n-1})$, we
we must retime each matching pair of syllables so that they are the
same length.  We first explain how we retime audio and video
syllables individually, and then describe our strategy for
combining these two methods to minimize visual and audible
artifacts.

\subsubsection{Audio retiming}
\label{subsubsec:audio_retiming}

Given an audio
syllable $a_i$ and a target length $L$, we speed up or
slow down the audio such that $\norm{a_i} = L$. 
We use Waveform Similarity Overlap-add (WSOLA)\,\cite{319366} as
implemented in Matlab by TSM-Toolbox\,\cite{DriedgerM14_TSM_DAFx} 
to retime each audio syllable. 
WSOLA produces a waveform by maximizing local similarity between the generated result and the original waveform in corresponding neighborhoods, as measured by short-time Fourier transform representations. %
We have also experimented with phase
vocoders\,\cite{flanagan1966phase}, Harmonic-Percussive
Separation\,\cite{6678724} and the commercial Elastique
algorithm\,\cite{elastique}, but found that WSOLA works best for our goal of retiming of short speech segments corresponding to syllables. Other methods excelled at retiming music or longer speech segments.

\subsubsection{Video retiming}
\label{subsubsec:video_retiming}

Given a visual syllable $v_i$ and a target length $L$, our goal is to
retime $v_i$ such that $\norm{v_i} = L$. Our approach is to treat the
original frames of the syllable as samples of a continuous
time-varying function parameterized over the time segment $[0, 1]$. We
then resample $L_f$ evenly spaced frames in $[0, 1]$, where $L_f$ is
the target length expressed as a number of frames (i.e. if $L$ is in
seconds, $L_f$ is set to $L$ times the video framerate). Since
resampled frames may not lie on frame boundaries we use either
nearest-neighbor sampling (for videos with rapid motion or low
frame-rate) or optical flow
interpolation\,\cite{10.1007/3-540-45103-X_50} to generate them. Also
note that we assume $L_f$ is an integer value, a property we enforce
in \Cref{subsubsec:combined_retiming}.

\subsubsection{Combined audio-video retiming}
\label{subsubsec:combined_retiming}

Retiming artifacts can appear as
speedup/slowdown in the audio or as blur/choppiness in the video. Such
artifacts become more prominent as the retiming factor
increases. Our approach is to prevent extreme retiming of either the
video or the audio by trading off retiming in one channel (audio or
video) for retiming in the other. 

Given a pair of matching syllable sequences $(a_1, \dots, a_n)$ and
$(v_j, \dots, v_{j+n-1})$ we define 
the sequence of lengths for each output syllable $(l_1, \dots, l_n)$.
Since the retiming factor is a multiplicative property (e.g. retiming 1 minute to 2 minutes will have similar quality to retiming 10 minutes to 20 minutes, and is very different from retiming 10 minutes to 11 minutes), we set $l_i$ to be the 
\emph{geometric} mean of $\norm{a_i}$ and $\norm{v_{j+i-1}}$,
\begin{equation}
l_i = \sqrt{\norm{a_i}\norm{v_{j+i-1}}}
\end{equation}
to evenly distribute retiming between audio and video.

Next we calculate the audio retiming factors for the sequence
\begin{equation}
f_i^\text{aud} = \frac{l_i}{\norm{a_i}}
\end{equation}
Extreme audio retiming produces audibly disturbing results.
Extreme video retiming is also undesirable, however we found that the visual slowdowns or speedups are often preferable to audio artifacts.
Thus, we limit the allowed audio retiming factor 
\begin{equation}
\hat{f}_i^\text{aud} = \max(\min(f_i^\text{aud}, \frac{1}{T}), T)
\end{equation}
Where $T$ is an audio retiming threshold parameter set empirically to 1.3. 
We avoid abrupt timing changes between consecutive syllables by convolving the sequence of retiming factors $(\hat{f}_1^\text{aud}, \dots, \hat{f}_n^\text{aud})$ with a box filter of size 3. 
The video retime factors $f_i^\text{vid}$ are set to produce the proper combined retiming amount
\begin{equation}
f_i^\text{vid} * \hat{f}_i^\text{aud} = \frac{\norm{a_i}}{\norm{v_{j+i-1}}}
\end{equation}
As a last step, we slightly update the retiming factors to the closest values that yield integer values for $L_f$. 
Given the audio and video retiming factor, we apply the methods in \Cref{subsubsec:audio_retiming} and \Cref{subsubsec:video_retiming} to produce the final result.

% ---------------------------------------------------------------------------

\section{Results}
\label{sec:results}

\begin{figure*}[ht]
  \setlength{\w}{\textwidth}
  
  \centering
  
  \includegraphics[width=\w]{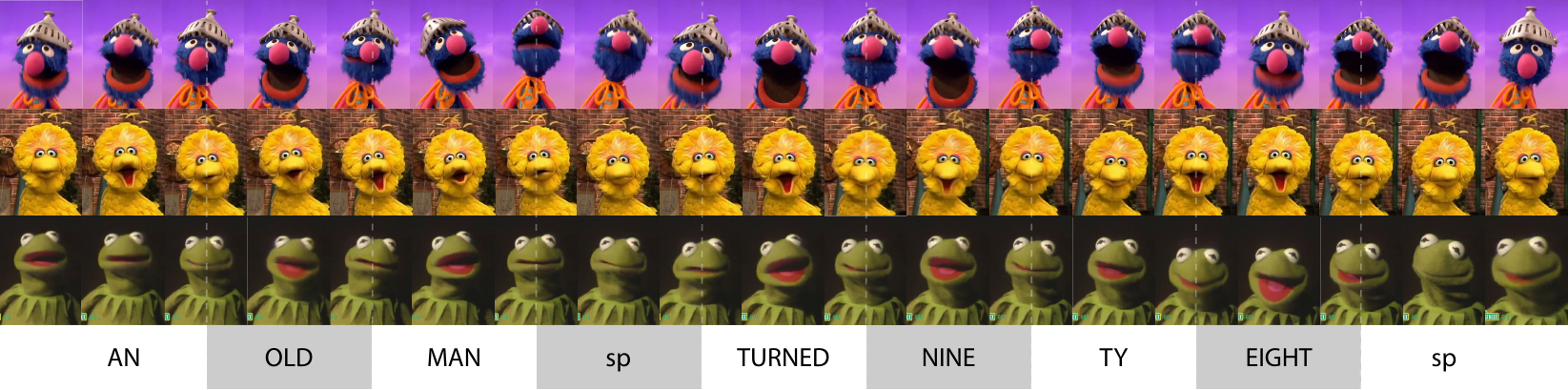}
  
  \caption{Given an audio recording and a puppet video, we find the
    best matching subsequence of the video and produce a retimed
    result that matches the audio. We show selected frames after our
    alignment using Kermit, Grover and Big Bird videos. The frames shown
    are from the start, midpoint and end of each syllable in the new speech (bottom).
    Notice how the
    beginning and end of each audible syllable generally corresponds
    with a closed mouth, and the middle with an open mouth, while
    silent syllables correspond with a sequence of closed mouth
    frames.  \textit{Full videos in supplemental materials.} 
}
  \label{fig:result_single_puppet}
\end{figure*}

Our main motivating application is the creation of puppet videos
dubbed with new speech content (\Cref{fig:result_single_puppet}). Our method can also be used to improve
synchronization in existing puppetry videos and to facilitate video
translation. 
We collected 9 puppet videos of Kermit (2x), Big Bird (2x), Grover, Miss Piggy, Fozzie Bear, Cookie Monster, Abby Cadabby and one Japanese anime video, ranging in length from 14 seconds to 68 seconds with an average length of 35 seconds. We also collected 55 audio snippets from the internet containing spoken text, movie quotes, jokes, songs and political speech, ranging in length from 2 seconds to 54 seconds with an average length of 8 seconds. 
We have generated all combinations of results mentioned in the paper (appearance-based/audio-based, with/without manual correction of audio annotation). For representative results with the various puppets, see [supplemental materials: Representative Results]. %

\Cref{fig:result_single_puppet} shows results of video extraction and
retiming. Our tool finds a well aligned video sub-sequence
(\Cref{sec:alignment}), retimes it (\Cref{sec:retiming}), and produces a
final result synchronized to the user's audio recording. 
We encourage the reader to look at the supplementary materials for video results. %
Producing a result using a 1 minute video clip with a 10 seconds audio snippet takes less than 10 seconds for nearest-neighbor sampling of frames, and 3 minutes for optical flow based interpolation on a MacBook Pro with a 2.7 GHz Intel Core i7 processor. We used nearest-neighbor sampling for Cookie Monster, as he moves erratically, causing adjacent frames to be dissimilar and unsuitable for optical-flow based interpolation. All other videos use optical flow. 

For comparison between fully annotated visual syllables and 10\% annotated visual syllables, see [supplemental materials: 10\% vs. 100\% annotation]. Many 10\% results are comparable to the fully annotated version, with errors arising when frames are mislabeled. A mislabeled \textit{open-mouth} may introduce an erroneous \textit{nsp} syllable, which causes a spoken syllable for a closed mouth sequence. A mislabeled \textit{closed-mouth} may split \textit{nsp} in two or change it to \textit{sp}. 

For comparison between appearance-based and audio-based results, see
[supplemental materials: appearance vs. audio]. Appearance-based
results are generally better than audio-based, as evident by our user
study (Section~\ref{sec:userstudy}). Our experience in examining the
original puppet videos is that puppeteers often do not perfectly align
mouth closed-open-closed sequences with the syllables in their
speech. The mouth articulations often lag the speech and these
misalignments create artifacts for our audio-based approach.

To evaluate the contribution of good audio annotations, see
[supplemental materials: Appearance-based, with/without manual correction of phoneme alignment]. We compare
between using the automatically generated phoneme alignment
(P2FA~\cite{yuan2008,rubin2013}) and a manually corrected alignment.
We find that P2FA produces good alignments at the level of words, but
is not perfect at the level of phonemes or syllables. These misalignments
appear as extra or missing visual syllables.

We also compare against baseline approaches ([supplemental
  materials:baselines]). Baselines include random selection of clips
and an ablation study in which we omit retiming (\Cref{sec:retiming})
from our algorithm.
The baseline methods are often completely
misaligned with the new speech as mouth movements occur when there is
silence in the speech and vice versa. 
In the no-retiming case our approach
simply aligned the audio and visual syllables, so the first syllable in the speech start with a mouth opening, but since there is no re-timing the video and audio quickly fall out of synchronization with each other.

As far as we know the proposed method is the first to dub puppet videos to new audio, and has no direct comparison in previous literature. Methods such as Video Puppetry \cite{barnes2008} require a puppeteer to explicitly perform, while our method produces the performance from the input audio sequence. Methods such as JALI \cite{edwards2016} require a rigged CG model, while we operate on internet videos. Methods such as Deep Video Portraits \cite{kim2018DeepVideo} require a driving video to control the result. Lastly, the method of Suwajanakorn et al. \cite{suwajanakorn2017} solves the arguably harder task of puppeteering human heads from audio, but requires hours of training video. In contrast, we only require one video that can be a few seconds long.

\vspace{0.05in}
\noindent
\textbf{\emph{Other applications.}}
    In addition to dubbing a puppet video with new audio, 
our methods can also be used to
improve existing puppet performances and for dubbing a performace in a different language. 
We
show an example from Sesame Street in which the audio and puppet mouth
are not well synchronized [supplemental materials: improve synchronization
  in existing puppet performances]. Interestingly, we found that such
misalignments in the original performance  often arise, even in professional productions. We can improve
the synchronization of the original performance (at a cost of retiming artifacts) by applying our algorithm on the original
video and audio content. Thus, our methods can be used as part of a
post-production pipeline to improve puppet performances.

\subsection{Algorithmic Evaluation}
\label{sec:algeval}
One goal of our method (\Cref{sec:methods}) is to minimize extreme
speed-ups and slow-downs due to large differences in syllable
lengths between the aligned audio and video. We can directly measure
these artifacts by calculating the average 
length ratio between audio and visual syllables (\Cref{tbl:alignment_eval}).
We compare between our alignment procedure applied to appearance-based
syllables and audio-based syllables.  
As a baseline, we also compute a random alignment where we randomly select a sequence of appearance-based syllables in the video with the same number of syllables as in the new speech.
We obtain less squash/stretch distortion compared to the baseline,
even though our alignment
has other constraints -- not just amount of stretch and squash, but
also the fact the silent and non silent regions should align.
Our appearance-based method outperforms the audio-based method in stretch/squash minimization.

\begin{table}

  \newcommand{\linesc}[1]{\begin{tabular}{@{}c@{}}#1\end{tabular}}
  \newcommand{\linesl}[1]{\begin{tabular}{@{}l@{}}#1\end{tabular}}
  
  \centering
  \begin{tabular}{@{}lccc}
    \toprule
                                             & Random & \linesc{Ours\\(audio)}  & \linesc{Ours\\(appearance)} \\
    \midrule
     \linesl{Mean length ratio\\(aud > vid)} & 3.70   & \linesc{4.25\\(n=612)}  & \linesc{2.92\\(n=590)}      \\[0.3cm]
     \linesl{Mean length ratio\\(vid > aud)} & 3.13   & \linesc{2.32\\(n=615)}  & \linesc{2.27\\(n=568)}      \\
    \midrule
     \linesl{Total}                          & 6.83   & \linesc{6.57}           & \linesc{\textbf{5.19}}      \\  
    \bottomrule
  \end{tabular}
  \caption{
  Average distortion amount per syllable. We calculate the average ratio between audio and video syllable length.
  Large values cause speedup/slowdown in the audio and blur/choppiness in the video. We out-perform a random selection of a video starting point, even though such random selection does \textit{not} need to accommodate other constraints, such as matching silence and non-silence respectively.}
  \label{tbl:alignment_eval}
\end{table}

\subsection{User Study}
\label{sec:userstudy}
\begin{figure}[t]
  \setlength{\w}{\columnwidth}
  
  \centering
  
  \includegraphics[width=\w]{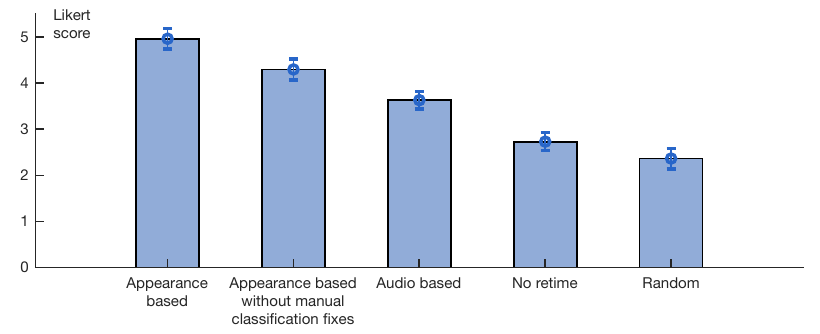}
 \caption{
 User study ratings. Our participants rated the appearance-based method with and without manual correction, audio-based method, results without retiming and a random selection of video segments on a 1--7 Likert scale.
 Bars indicate mean ratings, whiskers are standard error of the mean.
  }
  \label{fig:user_study}
\end{figure}

We conduct a user study to evaluate and compare our results. We evaluate the following conditions:
\begin{enumerate}
  \item \textbf{\emph{Appearance 10\% + Correction.}} Our appearance-based method with 10\% annotation and manual correction of wrong labels. 
  \item \textbf{\emph{Appearance 10\%.}} Our appearance-based method with 10\% annotation, without manual correction of wrong frame labels. 
  \item \textbf{\emph{Audio.}} Our audio-based method.
  \item \textbf{\emph{No-retiming.}} Our appearance-based method with 10\% annotation but without retiming. 
  \item \textbf{\emph{Random.}} Random selection of a video subsequence to match with the new audio.

\end{enumerate}
We use 5 audio-video pairs and generate the 5
conditions for each, resulting in 25
videos. We asked 6 people to rate each result on a 7-point Likert scale, ranging
from 1-extremely bad quality to 7-extremely good quality, for a total
of 150 ratings. The participants were free to view the videos in any
order and to view a video multiple times.
\Cref{fig:user_study} shows the study results. 
A Kruskal-Wallis test finds that there is a significant difference between these conditions ($p < 10^{-8}$). We then use Tukey's honestly significant difference procedure 
for pairwise comparisons and find that all pairs of conditions except 1-2, 2-3, 3-4 and 4-5 are significantly different ($p << 0.02$).
These results suggest that our appearance-based methods out-perform the other methods, followed by the audio-based method, our method with no retiming and a random baseline.

\subsection{Discussion}
\label{sec:discussion}

Together, our results and evaluations suggest that our appearance-based method
generally produces more convincing dubs than our audio-based approach
and that both outperform baseline dubbing methods. However our
appearance-based method does incur some manual annotation cost as
reported in Table~\ref{tbl:comparison}. Thus, the two methods offer a
tradeoff between visual quality and annotation time.  We recommend
using the appearance-based approach when accuracy is most important
and users can spend some effort on annotation. We recommend the
audio-based approach when users have no time to annotate the video --
e.g. if they need to quickly dub a large number of videos. 
\section{Future Work and Conclusion}
\label{sec:limitations}

Our approach has several limitations, suggesting interesting
directions for future work. 

\vspace{0.05in}
\noindent
\textbf{\emph{Automatic puppet mouth state detection.}}
Our appearance-based method requires some manual annotation
effort. One direction for future work is to collect a larger set of
puppet videos, and use them to train a generic puppet mouth state
detectors that can perform well for any unseen puppet videos.  Our
audio-based method relies on good alignment between the syllables in
the original audio track and the puppet's closed-open-closed mouth
states. While we have found in practice that most puppet videos are
not perfectly aligned, there may be enough alignment to serve as weak
supervision to build a puppet mouth state detector without any manual
labeling.

\vspace{0.05in}
\noindent
\textbf{\emph{Background and camera motion.}} Our current method does not analyze the background behind the puppet. While we found this to work for most puppet videos, some videos contain regular background motion. For such videos the retiming procedure will produce irregular motion, which can be visually disturbing (e.g. a smooth camera pan can become choppy). Adding background analysis to the matching procedure will alleviate such artifacts, as a cost of more computation.

\vspace{0.05in}
\noindent
\textbf{\emph{Generalization to cartoonized human faces.}}  Our work is
puppet specific and relies on guidelines from expert puppeteers on how
to give convincing performances. It is not designed to work on human
faces, which are far more expressive than puppets ([supplemental
  materials: human heads]). We would like to
investigate this space between our method and existing human speech
manipulation methods \cite{bregler1997,suwajanakorn2017}. For example,
some animated characters rely on mouth openness, but also on mouth
contents (teeth visibility) and gestural appendage movements to more
expressively convey speech
and emotion. We plan to investigate these and other examples in the
under-explored spectrum between puppets and humans.

\vspace{0.05in}
\noindent
Despite these limitations our tools offer new ways to quickly create
dubbed puppet videos. As the amount of video available online
increases we believe that such remixing techniques will become more
commonplace as they offer efficient approaches to produce high-quality
visual stories.

\bibliographystyle{eg-alpha-doi}
\bibliography{main}

\end{document}